# Optical pulse induced ultrafast antiferrodistortive transition in SrTiO$_3$

Saqeeb Adnan[1], Amey Khanolkar[2], Shuxiang Zhou[2], David H. Hurley[2], Marat Khafizov[1‡]

[1] Department of Mechanical and Aerospace Engineering, The Ohio State University, Columbus, OH 43210, USA

[2] Idaho National Laboratory, Idaho Falls, ID 84315, USA

## ABSTRACT

The ultrafast dynamics of the antiferrodistortive (AFD) phase transition in perovskite SrTiO$_3$ is monitored via time-domain Brillouin scattering. Using femtosecond optical pulses, we induce a thermally driven tetragonal-to-cubic structural transformation and detect notable changes in the frequency of Brillouin oscillations (BO) induced by propagating acoustic phonons. First, we establish a fingerprint frequency of different regions across the temperature phase diagram of the AFD transition characterized by tetragonal and cubic phases in the low and high temperature sides, respectively. Then, we demonstrate that in a sample nominally kept in tetragonal phase, deposition of sufficient thermal energy induces an instantaneous transformation of the heat-affected region to the cubic phase. Coupling the measured depth-resolved BO frequency with a time and depth-resolved heat diffusion model, we detect a reverse cubic-to-tetragonal phase transformation occurring on a time scale of hundreds of picoseconds. We attribute this ultrafast phase transformation in the perovskite to a structural resemblance between atomic displacements of the R-point soft optic mode of the cubic phase and the tetragonal phase, both characterized by anti-phase rotation of oxygen octahedra. Evidence of such a fast structural transition in perovskites can open up new avenues in the field of information processing and energy storage.

## 1. Introduction

By enabling up to terahertz speed signal modulation in nanodevices, ultrafast phase transitions have the potential to transform data processing, transmission, and storage technologies[1,2]. The primary idea involves exploiting the tunable structural[3], magnetic[4], and electrical[5] properties of material susceptible to ultrafast changes in crystal structure or electronic configuration. For instance, the temperature-induced insulator/conductor switching in vanadium dioxide ($VO_2$) is incorporated in the design of ultrafast transistors that operate at GHz frequencies[6]. In addition, novel storage devices such as phase-change memory (PCM), which can store data at high speeds and densities, are being developed utilizing the ultrafast reversible transition between amorphous and crystalline structure in $Ge_2Sb_2Te_5$[7]. The development of these potential technologies demands a deeper understanding of the nature and dynamics of the phase transition. Consequently, there has been extensive research focused on the phase transition relating to electronic and/or magnetic degrees of freedom[8–11]. Through ultrafast laser and x-ray diffraction techniques, the rate and evolution of these transitions have been determined[12–14]. While there have been breakthroughs with electronically driven phase transitions (such as those associated with electron spin reorientation[15] or symmetry changes of valence charge/orbital number[16]), the timescales of purely structural phase transitions are yet to be carefully explored. In this study, we investigate the laser-induced ultrafast structural phase transition in strontium titanate (STO), a perovskite oxide.

Due to their unique structure and functional properties, perovskites have received substantial attention from a scientific standpoint over the past few decades[17]. Remarkable technological breakthroughs have been made by harnessing their ability to undergo reversible phase transitions induced by external stimuli, such as temperature, pressure, or electric fields. STO is an archetypical perovskite and has garnered significant interest for its potential application in energy conversion processes, particularly in the generation of thermoelectric power[18], photocatalytic water splitting for hydrogen production[19], as oxygen sensors[20], and as anodes in fuel cells[21]. The electrical conductivity of STO can be easily tailored from insulating to metallic regimes by substitutional doping with lanthanum or niobium ions, making it an attractive substrate for epitaxial growth of high-temperature superconducting and other functional material thin films[22–24]. It has also been proven useful in the development of ultrafast nanotransistors[25,26].

STO undergoes an antiferrodistortive (AFD) second-order structural phase transition at $T_{AFD}$ ~105 K, where the crystal symmetry lowers from cubic to tetragonal (space group $I4/mcm$) upon cooling. This structural modulation manifests itself by an antiphase rotation of oxygen octahedra ($TiO_6$) of adjacent unit cells within the basal plane around the Ti atom[27,28], resulting in a tetragonal distortion. In the cubic phase, STO exhibits paraelectric behavior and continues to be paraelectric at low temperatures across the structural phase transition despite a very high dielectric constant. The distortion of the oxygen octahedra causes anti-polar displacements of the Sr-site, which in turn, suppresses the ferroelectric instability (FE)[29]. The FE phase of STO is too weak to be revealed at low temperatures due to competing interactions from quantum fluctuations and antiferrodistortion and thus remains hidden[30]. However, in a recent study by Li et. al., the hidden FE phase was dynamically induced using THz electric field excitation and the transient phase shift was on the order of ten picoseconds[31]. On the other hand, the antiferrodistortive (AFD) phase transition in STO has been the subject of extensive theoretical and experimental research over the past six decades and has provided critical insight into the dynamic relationship between the AFD instabilities stemming from $TiO_6$ octahedral rotation and the soft optic modes near the Brillouin zone edge[28,32–35]. While there have been a few previous investigations aimed at uncovering evidence of ultrafast rotation of the octahedral in perovskite oxides[36,37] using time-resolved x-ray diffraction experiments, these studies have focused only on the ultrafast transient increase of reflection intensities related to the rotation of the oxygen octahedra following photoexcitation above the band gap. The experimentally time-resolved dynamics of a structural transformation between a cubic to a tetragonal phase as a function of temperature, however, have not been systematically studied.

Herein, we present the first experimental evidence of the picosecond timescale AFD structural phase transition in STO using ultrafast laser excitation. The transient heat deposition from the intensity-modulated pump pulse train drives the cold tetragonal crystal into the cubic phase. We utilize the time-domain Brillouin scattering (TDBS)[38] technique to image the time- and depth-resolved phase profile of the STO sample by tracking the frequency of the coherent longitudinal acoustic (LA) phonon mode. We observe that the phase transition is accompanied by a damping of the LA vibrational mode near the Brillouin zone boundary. The increased dissipation near the phase transition has been attributed to the coupling of acoustic phonons to the soft optic mode which leads to an anomaly in the elastic constants at the phase transition[34]. Through careful analysis of the time-resolved Brillouin frequency plots coupled with a heat diffusion model, we determine the timescale of the transient AFD transition across the sample's depth to be on the order of a hundred picoseconds. A structural transition at such ultrafast timescales offers opportunities for ultrafast tuning of the electromagnetic properties that can lead to novel switching and phase change devices.

## 2. Results and Discussion

Figure 1a shows the transient reflectivity signal of a probe beam in gold-coated STO following excitation by a modulated infrared pulse acting as a pump launching an acoustic wave. The transient signal comprises Brillouin oscillations (BO) superimposed on top of the electronic and thermal background response which are responsible for the initial abrupt rise. These oscillations originate from the constructive/destructive interference between two reflections of the probe, one from the sample surface and the other from the localized strain field of the ultrasonic wave propagating in the bulk[38]. The frequency of the Brillouin oscillations under a collinear and normal incidence of the pump and probe is given by[39]:

$$f = \frac{2nv}{\lambda},\qquad(1)$$

where $f$ is Brillouin frequency, $n = 2.567$ index of refraction at $\lambda = 395$ nm probe beam wavelength[40], and $v$ is velocity of ultrasonic wave. This configuration tracks one-dimensional propagation of the ultrasonic waves normal to the sample surface. Any depth dependent microstructural information is encoded in the BO frequency[38,41,42].

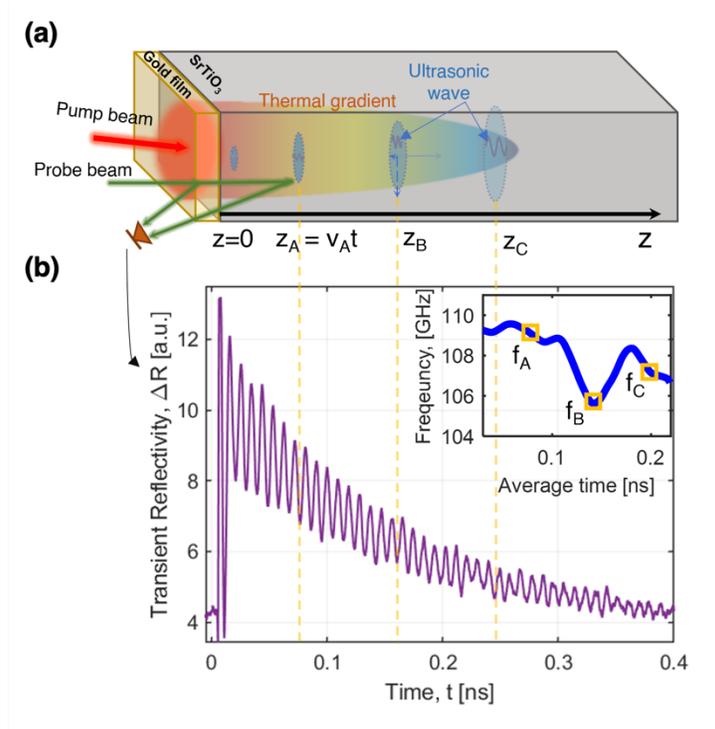

**Figure 1.** a) Schematics for the TDBS measurements. The thermoelastic excitation from the pump pulse generates ultrasonic waves that propagate along the sample's depth normal to its surface. The probe beam generates two reflections (green arrows): one from the sample surface and another from the strain field of the propagating ultrasonic pulse signal, represented by the blue disc. These reflections create an interference pattern in the time-resolved reflectivity change ($\Delta R$) of the probe beam, with a frequency and period proportional to the ultrasonic velocity and the index of refraction of the probe beam in the material. The reflection angles are normal to the surface but are shown at an angle to demonstrate two reflections. The signal at a given time

reveals microstructural information at the location of the ultrasonic wave. b) Time-resolved reflectivity change from the TDBS measurements on cubic STO single crystal at 86 K. The Brillouin oscillations, superimposed on top of the electronic or thermal background, are isolated using a 7th-order polynomial. The frequency of the full temporal signal gives an average of the material response from the probed region along its depth. However, localized responses are also achievable by obtaining frequency across a narrow window of the transient signal. The inset shows different BO frequencies obtained in a single transient signal indicating varying material response along the depth of the sample.

Figure 2a shows the temperature-dependent BO frequency measured under different pump beam fluences. The frequencies were obtained by fitting the full temporal scan to a damped sinusoid[38]:

$$A \sin(2\pi f t - \varphi) \exp\left(-\frac{t}{\tau}\right) \qquad (2)$$

where $f$ is the frequency, $\tau$ is a damping time constant of the oscillation, and $\varphi$ is a phase offset. The reported temperature is the ambient temperature recorded by a thermocouple ($T_o$) attached to a location far removed from the interrogation region of the sample. BO frequencies reported in Figure 2, representing an average across a 2 $\mu$m thick layer, follow closely the temperature trend of the elastic constants as STO undergoes cubic to tetragonal phase transition with a notable minima at the phase transition temperature [43,44]. However, owing to the localized heating proportional to the pump fluence, the apparent phase transition temperature $T_{PH}$ shifts to a lower temperature. Based on Figure 2, we can define the characteristic fingerprint frequency of each point in proximity of the cubic-to-tetragonal transition of the phase diagram. The tetragonal phase is characterized by 108 GHz, the cubic phase has a fingerprint frequency of 110 GHz, and the actual transition is represented by BO frequencies lower than 107 GHz.

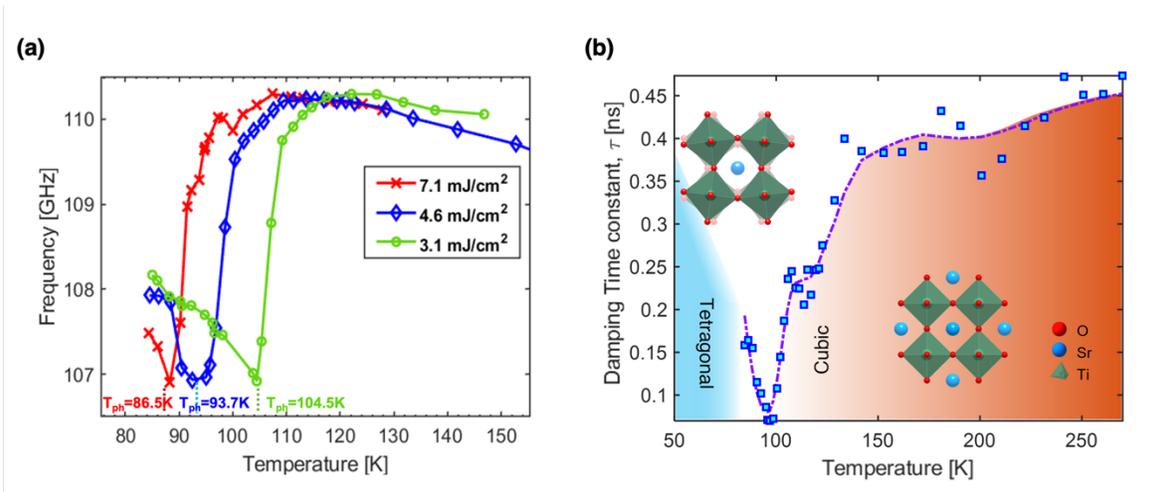

**Figure 2.** Frequency and damping time constant versus thermocouple temperature. a) High laser fluence-induced shift in $T_{ph}$ demonstrated by temperature-dependent Brillouin frequency profiles. b) Damping of the transient reflectivity oscillation near $T_{ph}$. The damping time constant is obtained by fitting the transient reflectivity profile for each temperature using equation (2). The sharp attenuation of the time constant separates the cubic and tetragonal crystal phases of STO as illustrated.

We found significant attenuation of the BO near the phase transformation temperature (see supplementary document). The damping time constant ($\tau$) obtained by fitting the transient signal using equation (2) is plotted in Figure 2b. This is consistent with previous observations involving inelastic neutron scattering and Raman spectroscopy measurements[27,32,43,45,46]. Softening of these phonon modes in the tetragonal phase near $T_{AFD}$ has also been reported, consistent with our measurements[47]. The origin could be attributed to either the short lifetime of the acoustic phonons[48,49] or the formation of the ferroelectric domain walls whose boundaries undergo fluctuations on the time scale of the experiment[50]. Experimental studies have shown that this softening of phonons near transition temperature is related to the coupling between the lattice and electronic degrees of freedom[51].

In Figure 3, we present time-resolved BO frequency, obtained by evaluating the frequency across a narrow time window (~0.045 ns, corresponding to 5 oscillations), which provides localized information on the ultrasonic wave velocity as a function of depth[41,52]. Selected temperatures are plotted in Figures 3a-e (transient reflectivity data for complete temperature set is available in the supplementary document). Temperature-dependent and time-resolved BO reveal interesting dynamics. We observe that BO frequency varies as a function of depth and follows a trend that suggests that acoustic wave is probing a different STO phase as it propagates deeper into the sample. When the sample is kept above or well below $T_{AFD}$, the frequency remains constant, which indicates that the same phase is probed across different depths (Figures 3a and e). However, when the sample is kept just a few degrees below the $T_{AFD}$, the BO frequency varies as a function of depth (Figures 3b-d). At early time delay, the frequency is about 110 GHz and matches the BO frequency of the cubic phase. At longer time delays, it is 108 GHz and matches the BO frequency of the tetragonal phase. At the intermediate time delay, the frequency drops down to 106 GHz which corresponds to a phase transition region. Interestingly, the location of the frequency dip occurs at later delay times as the ambient temperature approaches $T_{AFD}$. The trends observed in Figure 3 serve as an illustrative demonstration of a selective phase tuning of STO by an optical beam when the sample's temperature is maintained a few degrees below $T_{AFD}$. Next, we analyze these observations considering that a temperature gradient is established upon absorption of the pump laser beam by the thin transducer gold film and is responsible for the observed switching of the surface of the nominally tetragonal sample into the cubic phase.

Heat is deposited at the surface upon absorption of the pump pulse and gets dissipated into the bulk of the sample. The surface heating causes the temperature of the regions closer to the surface to reach temperatures above $T_{AFD}$ while the deeper regions remain cold. This is evident in Figures 3b-d, where we see a distinct dip in the frequency (assigned as the AFD transition point), that separates the heated 'cubic' region and the remaining colder segment underneath that retains the tetragonal phase. The fingerprint frequency of different phases allows us to identify a depth $z_{AFD}$ at which the heated region switches back to the tetragonal phase. In Figure 3b-d, we see that the thickness of the transformed layer becomes larger (the location of the dip occurs at a longer delay time) as we increase the ambient temperature. In other words, with the sample surface getting more intensely heated, the AFD transition had to occur at a greater depth as evident in our measurement. This led to the conclusion that the localized phase transition occurring within different depths of the sample is primarily driven by temperature rise due to pump heating. A similar conclusion can be made when transient frequency profiles at similar temperatures but different fluence are analyzed (supplementary document). At $T_o$ = 95 K, the phase transformation back to tetragonal occurs at $t$ = 0.21 ns at 7.1 mJ/cm$^2$ fluence (Figure 3d), whereas for the reduced laser fluence of 4.6 mJ/cm$^2$, this transformation occurs much earlier at $t$ = 0.09 ns (Figure S5b), i.e., smaller transformed thickness at lower deposited energy.

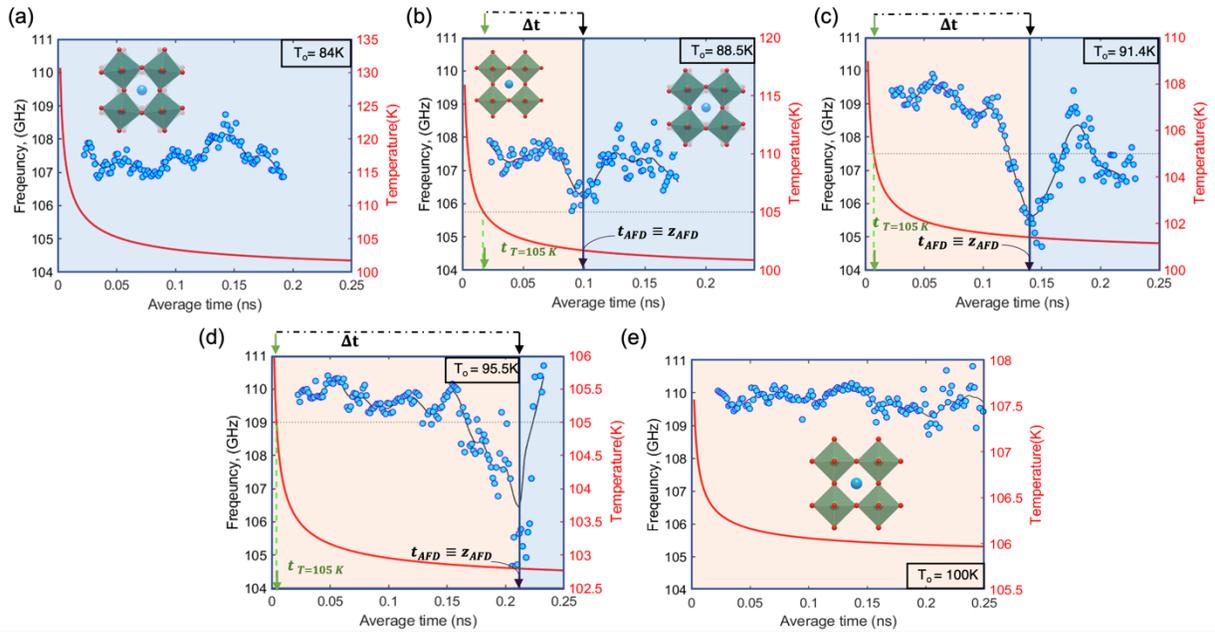

**Figure 3.** Evidence of ultrafast AFD phase transition captured within individual transient reflectivity signals. (a-e) The evolution of the transition with increasing ambient temperature measured by the thermocouple ($T_o$) is illustrated with transient frequency profiles for the case of 7.1 mJ/cm$^2$ laser pulse. The blue markers are moving window average frequency values obtained by fitting a short window capturing 5 oscillations of the transient reflectivity signal using equation (1). The solid black line is the fitted curve to the frequency data using weighted linear least squares and a 2nd order polynomial model with reduced weight to outliers. Different regions are color-coded by blue (tetragonal) and orange (cubic) shades based on the frequency fingerprint of different phases shown in Figure 2a indicating the crystal structure of the sample at various depths. The transition point is characterized by the sharp dip in frequency at $t_{\mathrm{AFD}}$, which moves further to the right with increasing ambient temperature. For plots (b-d), the transient temperature profile at the locations of these transitions ($z_{\mathrm{AFD}}$) is calculated using the thermal wave model and is shown by the solid red line. These lines are overlayed to obtain the structural relaxation time, Δt. The horizontal dotted line denotes $T_{AFD}$ = 105 K. Aided by the transient temperature profile at $z_{\mathrm{AFD}}$, the instance at which the temperature of that location reaches 105 K can be identified (green arrow). Whereas, the black arrow indicates when the transition becomes evident in the transient signal, $t_{\mathrm{AFD}}$. The difference between these two instances is defined as the structural relaxation time, Δt. For plots (a) and (e), the transient temperature profile is calculated at the highest (hottest) and lowest (coldest) points in the spatial resolution, respectively, in order to show the existence of only a single phase. In the case of ambient temperature, $T_o$= 84 K, the very large temperature rise on the sample surface did not result in the detection of the cubic phase due to the structural time lag. On the other hand, for $T_o$ = 100K, even the coldest section of the sample that can be detected has a temperature profile that never goes below 105K, indicating a complete transition to the cubic phase.

To gain a more accurate understanding of the timescale of the phase transitions, we performed a quantitative analysis of the laser-induced depth and time-resolved temperature changes across the region probed by the propagating ultrasonic wave. We implement a two-dimensional axisymmetric thermal wave model to capture the thermal response of the sample [53,54]. We consider the transient temperature profile at a fixed depth, specifically $z_{\mathrm{AFD}}$, the location where the local BO frequency reveals the phase transition. These transient temperature profiles $T_{\mathrm{tran}}(t, z_{\mathrm{AFD}})$ plotted as solid red lines in Figures 3b-d provide insight into temperature evolution at this particular location. One can see the initial jump in temperature is sufficiently large to drive the cold tetragonal crystal (T<105K) at $z_{\mathrm{AFD}}$ location into the cubic phase. After the temperature drops below 105 K, the crystal doesn't immediately transform to a tetragonal phase at $t = t_{T=105\ \mathrm{K}}$. This originates from the fact that structural transition is not instantaneous, and it takes a finite amount of time for the atoms to rearrange themselves, known as the structural relaxation time [55]. To evaluate this structural relaxation time Δt, we calculate the difference between the instance when the temperature at $z_{\mathrm{AFD}}$ drops to 105 K ($t_{T=105\ \mathrm{K}}$) and the instance when the transition is evident in the BO frequency profile ($t_{AFD}$). The $\Delta t$ for different ambient temperatures near $T_{AFD}$ is indicated in Figures 3b-d and it is in the order of a few hundred picoseconds.

Figure 4a summarizes the trend in relaxation times ($\Delta t$) obtained at different ambient ($T_o$) temperatures under 7.6 and 4.6 mJ/cm$^2$ laser fluences. Here for the x-axis, we use a scaled temperature to account for steady temperature rise of the probed region, which allows us to compare measurements at two different fluence levels. This scaling is motivated by an attempt to experimentally capture the temperature difference between $T_{AFD}$ and the temperature at which dynamic transformation occurs for a set of ambient temperatures and laser fluence. The overlap of relaxation times between different fluences at scaled temperatures also provides a critical validation for the choice of parameters in our thermal model. We notice that the structural lag is much higher when the ambient temperature is increased. From a physical perspective, it is more instructive to look at the relaxation rate which exhibits temperature trends consistent with the structural order parameter within Landau theory of phase transition in STO[56,57]. The order parameter is conveniently represented in terms of the rotation angle of the oxygen octahedra[58].

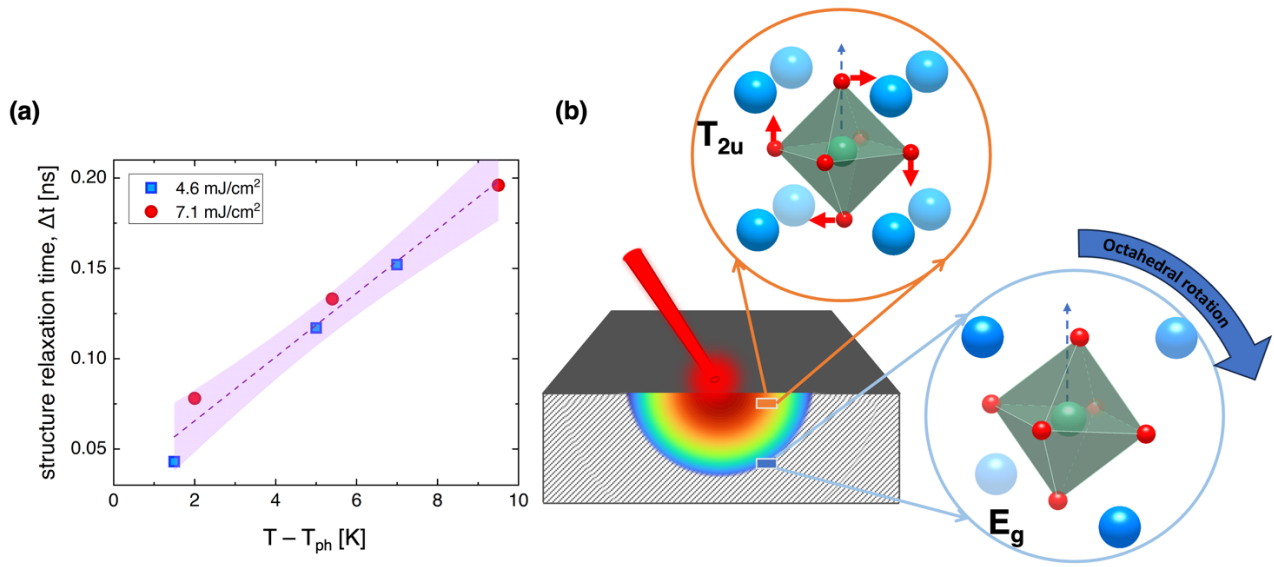

**Figure 4.** a) Increase in structural relaxation time away from the AFD phase transition temperature. The blue and red markers are the structural relaxation time using different pump laser fluences obtained from the transient frequency plots in Figures 3 and S5. The horizontal axis represents the temperature difference between the sample ambient temperature ($T_o$) during each measurement and the apparent transition temperature $T_{ph}$. The shaded region is the confidence band for the linear fit of the data. b) Different phonon vibrational modes across the sample depth (octahedral rotation angle is exaggerated). The hotter cubic region is representative of the triply-degenerate $T_{2u}$ ($\Gamma_{25}$) vibrational mode characterized by the planar rotation of oxygen octahedra around the third axis[27,59]. In contrast, in the cold tetragonal phase, this is split into either $A_{1g}$ or $E_g$ vibrational mode resulting from the tetragonal distortion.

It is remarkable to observe such a short relaxation time for a structural transformation. We postulate that such a fast structural relaxation is tied to an atomic structure resemblance between the soft R-point optical phonon mode with $T_{2u}$ symmetry and the tetragonal phase (Figure 4b). The cubic phase of STO (space group $Pm3/m$) is representative of the ideal cubic perovskite $ABO_3$ structure with a cubic arrangement of aligned $TiO_6$ octahedra (Figure 3e). The tetragonal phase (space group $I4/mcm$) is represented by the anti-phase rotation of the $TiO_6$ octahedra in the adjacent unit cells (Figure 3a). This atomic arrangement of the octahedra coincides with the atomic motion of the atoms in a $T_{2u}$ phonon mode (Figure 4b), which is the lowest phonon branch in the Brillouin zone boundary at the R-point of the cubic phase[27,60]. Recognizing this, we postulate that the relaxation time for the cubic-to-tetragonal transformation is proportional to the magnitude of the octahedron rotation away from its equilibrium position in the tetragonal phase. After the thermal impulse, the heat-affected zone recovers to its equilibrium state faster when the initial tetragonal-to-cubic transformation results in a larger octahedron rotation.

Ideally, one would expect that first principles atomic level calculations are used to support the aforementioned mechanisms. We would like to point out the difficulty of such analysis using existing standard density functional (DFT) based lattice dynamics models[61]. They work well in moderately anharmonic systems, where interatomic force constants (IFC) are obtained from static structure DFT calculations[62]. It has been reported that to capture IFC in STO correctly at finite temperatures one needs to perform a dynamic calculation using molecular dynamics[34]. Empirical interatomic potentials lack accuracy and researchers have just started to have a handle on performing self-consistent phonon calculations based on ab-initio MD[34,63]. Unfortunately, the high phonon anharmonicity of STO makes it difficult to assess the validity of this conclusion from first principles and remains a grand challenge.

Another possible explanation is that phase transformation dynamics is proportional to the magnitude of thermally induced strain. The current experimental configuration results in a larger thermal gradient near the sample surface compared to the small gradient beneath. With a higher ambient ($T_0$) temperature, the phase transition was occurring deeper into the sample (as evident in Figure 3) where the thermal gradient was weaker, causing a less favorable environment for the structural transition.

Due to the nature of the TDBS measurement technique, this phenomenon is evident only for a narrow window of ambient temperatures. At lower temperatures, even though the temperature of the sample is very high near the surface and is certainly in the cubic phase, we cannot detect it in our measurement (Figure 3a). This is because the probe delay for that small depth (~2 um) is very short compared to the structural relaxation time. On the other hand, in the case of higher ambient temperature (Figure 3e), we cannot probe deep enough with our setup to find regions with $T$<105 K.

In summary, we have utilized the time-domain Brillouin scattering technique to isolate the tetragonal-to-cubic structural transition. The experimental data and thermal wave model together demonstrate a laser-induced purely structural transition in the order of a few picoseconds. Experimental evidence of such ultrafast structural transition may prove to be pivotal in the advancement of nanoscale THz transmission technology.

### 3. Methods:

**TDBS setup**: Time domain Brillouin oscillation (TDBS) technique utilizes a femtosecond ultrasonic pump-probe laser setup to measure acoustic phonon response. Both the pump and probe beams were derived from the same ultrafast-mode-locked Ti: sapphire laser with a pulse width of 150 fs and a repetition rate of 80 MHz (Coherent Chameleon). The pump beam (790 nm) is modulated at 700 kHz using an acousto-optic modulator. The excitation of ultrasonic waves is achieved through the pump beam, while the probe (395 nm) beam gets photo-elastically coupled with the ultrasonic waves as they propagate through the sample depth. The small changes in the optical reflectivity of the probe beam are measured using lock-in detection (Stanford Research SRS854). A transient reflectivity change is measured by utilizing a mechanical stage that introduced a delay, t in the time-of-arrival between the pump and probe pulse. A 50x objective lens is used to focus both of the beams on the sample surface directed at a normal incidence angle. The GHz frequency oscillations (referred to as BO) originate from the interference between the reflected probe pulse from the sample surface and the strain field of the propagating ultrasonic wave as illustrated in Figure 1.

**Sample Preparation**: Single crystal SrTiO$_3$ samples, with surfaces oriented along the (100) crystallographic directions, were acquired from MTI Corporation. The SrTiO$_3$ crystals were mechanically polished and a thin (~ 7 nm) film of gold was sputter-coated on the surface. Since SrTiO$_3$ is optically opaque, the gold film acts as a transducer layer that absorbs the pump laser pulse energy and generates coherent phonons that propagate into the bulk. STO sample was mounted on a copper mount using a thermal plate inside liquid-nitrogen cooled in an optical cryostat (Cryo Industries model XEM). The cryostat chamber was cyclically purged using ultrahigh purity nitrogen, followed by continuous pumping to maintain pressures below 1 mTorr. TDBS measurements were conducted in a homogeneous region of the sample over a 79 K to 300 K temperature range. To improve accuracy, the temperature recorded by the thermal sensor was calibrated by an embedded thermocouple on the edge of the sample surface.

**Thermal Wave model**: By solving the 3D heat diffusion equation, this model describes temperature rise through the generation of thermal waves by a modulated heat source[53,54,64]

$$T(t, z, r = 0) = T_{ss}(z, f) + T_t(t, z, f_m, f_{rep}) = \qquad (3)$$

$$= \int_0^\infty A(\xi, f) e^{-\eta z} e^{-\frac{\xi^2 R^2}{2}} J_0(\xi r) \xi d\xi + e^{-i2\pi f_m} \sum_{n=-\infty}^{\infty} e^{-i2\pi(nf_{rep})t} \int_0^\infty A(\xi, f_m + nf_{rep}) e^{-\eta z} e^{-\frac{\xi^2 R^2}{2}} \xi d\xi$$

$$\text{where,} \quad A(\xi, f_m + nf_{rep}) = \frac{P}{4\pi\kappa} \frac{1}{\eta}; \quad \eta = \sqrt{\xi^2 + i2\pi(f_m + nf_{rep})/D}$$

Here, a steady state component, $T_{ss}$, which is entirely depth-dependent, and a transient part contributes to the total temperature rise. The transient temperature rise, $T_t$ is driven by the laser power ($P$), repetition rate ($f_{rep}$), modulation frequency ($f_m$). The laser spot size ($R$) was kept ~1$\mu m$ and the thermal conductivity ($\kappa$) and diffusivity ($D$) were obtained from previously reported values[65].

## Acknowledgments

This work was supported as part of the Laboratory Directed Research and Development Program at Idaho National Laboratory under the Department of Energy (DOE) Idaho Operations Office (an agency of the U.S. Government) Contract DE-AC07-05ID145142. D.H.H. acknowledges support from the Center for Thermal Energy Transport under Irradiation (TETI), an Energy Frontier Research Center (EFRC) supported by the DOE Office of Science Basic Energy Sciences (BES).

# Supplementary Material

**TDBS setup:**

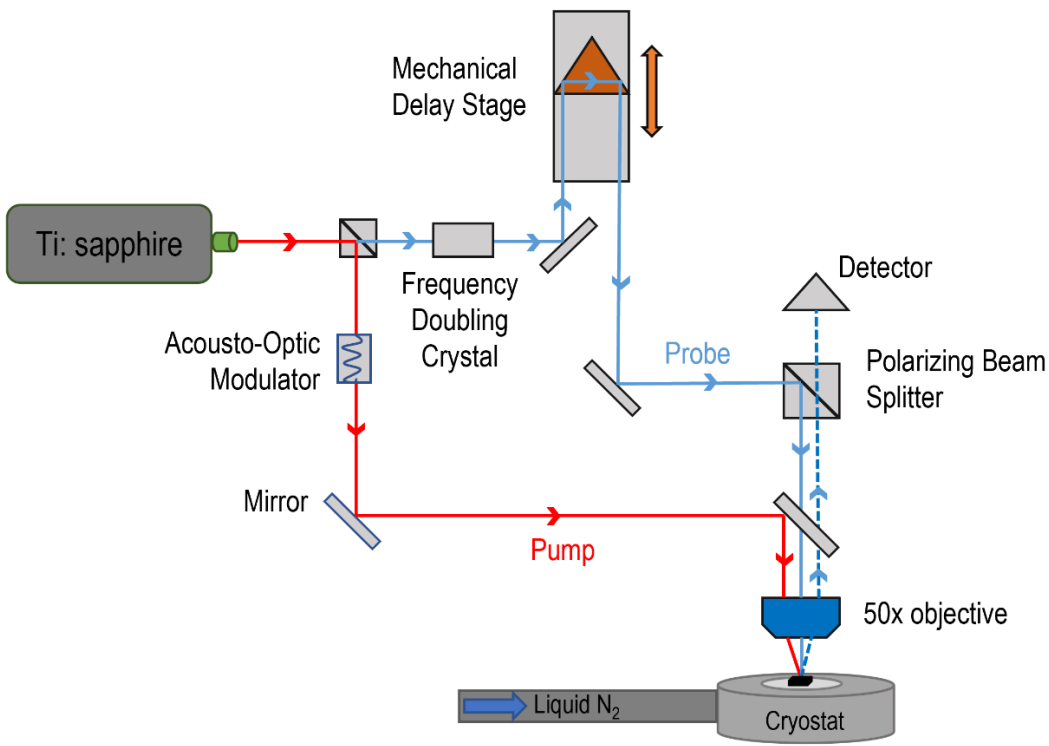

**Figure S1.** Schematic of Time-Domain Brillouin Scattering (TDBS) setup with the sample inside liquid N2 cooled optical cryostat. A Ti: sapphire laser with a pulse width of 150 fs and a repetition rate of 80 MHz produced both pump (790 nm) and probe (395 nm) beams. The pump beam was modulated at 700 KHz with an Acousto-optic modulator (AOM). A motorized stage was utilized to introduce a delay in the arrival of the probe beam with respect to the pump. Both the pump and probe beam were focused onto the sample using an optical cage and a 50x objective lens.

## Damping of the transient signal:

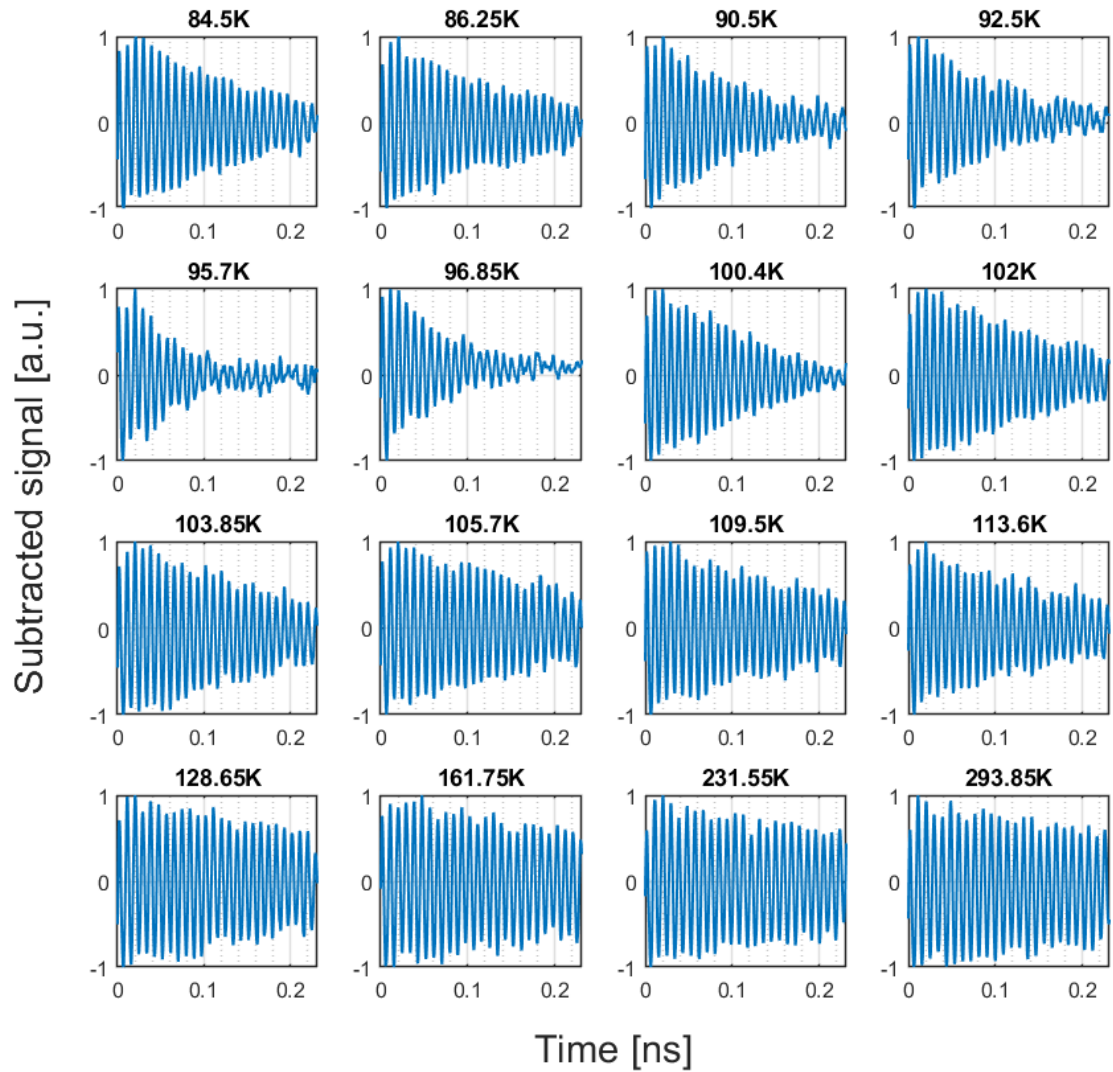

**Figure S2.** Damping of the transient reflectivity signals near $T_{ph}$ (=95.7K for 4.6 mJ/cm² laser fluence). The signals are background subtracted and normalized. The dampening of the signals is quantified using a damped sine wave and is illustrated in Figure. 2 of the main text.

## Transient Frequency Profiles:

Since TDBS is a depth profiling technique, different parts of the transient signal correspond to different regions in the sample. So, focusing on a narrow time window essentially provides information on a localized response. One can determine the BO frequency of a small window of the signal and move the window very slightly to obtain a response from deeper regions. Repetition of this process throughout the whole transient signal gives a complete depth-dependent frequency profile. For our analysis, we chose the length of the window to be ~0.045 ns and moved the window 0.0015 ns to the right after each acquisition of frequency. Each narrow segment of the transient signal was fitted with a damped sine wave to obtain the BO frequency. The transient frequency plots overlayed with temperature profiles are illustrated here.

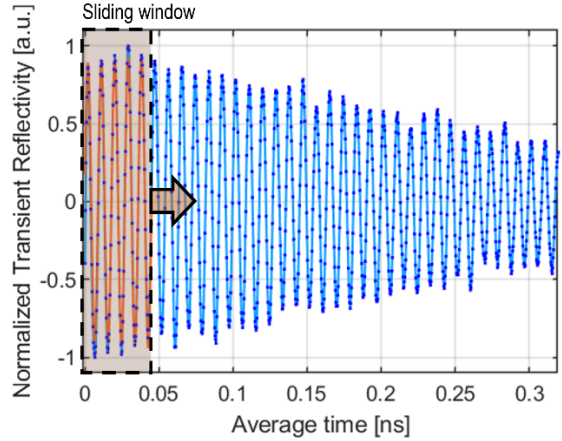

**Figure S3.** Localized frequency extraction using a narrow window of transient reflectivity signal. Each extraction window corresponds to one of the data points in the transient frequency plots.

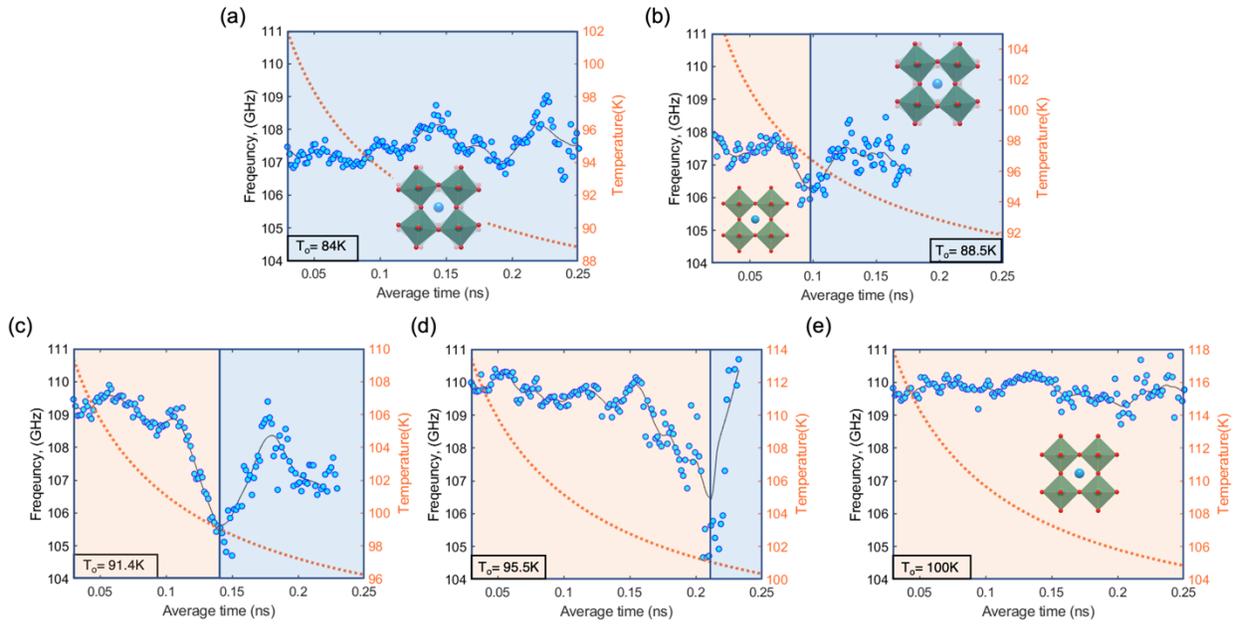

**Figure S4.** Transient frequency profiles for different ambient temperature ($T_o$) overlayed with instantaneous temperature profiles. The localized frequency values (blue markers) are for 7.1 mJ/cm² laser pulse (same dataset as figure 3). The dashed lines represent the instantaneous temperature of the sample at each probing depth, $T_{inst}(t, z = v_a t)$, which essentially captures the spatiotemporal temperature for all the experimental data points. It is evident that we do not detect the transition dip at the exact moment $T_{inst}$ reached 105 K, indicating the presence of structural time lag.

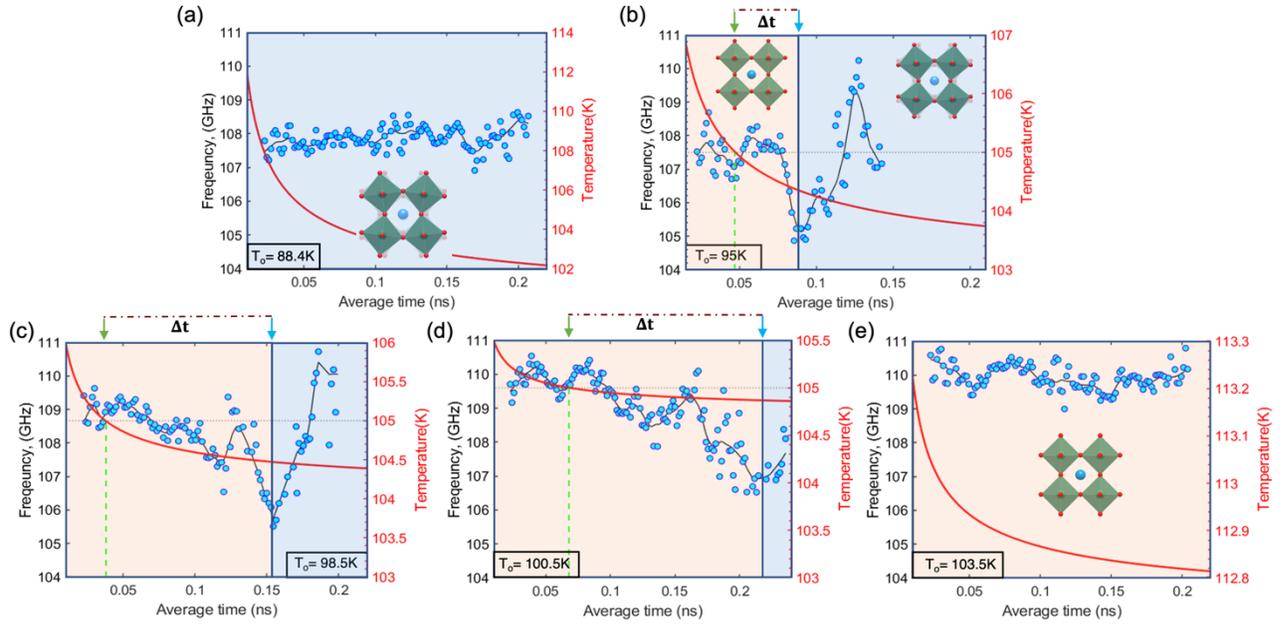

**Figure S5.** The evolution of the transition with increasing ambient temperature measured by the thermocouple ($T_o$) is illustrated with transient frequency profiles for the case of 4.6 mJ/cm² laser pulse. Similar to Figure 3 in the main text, evidence of transient AFD transition in ultrafast timescale with a different laser fluence is captured.

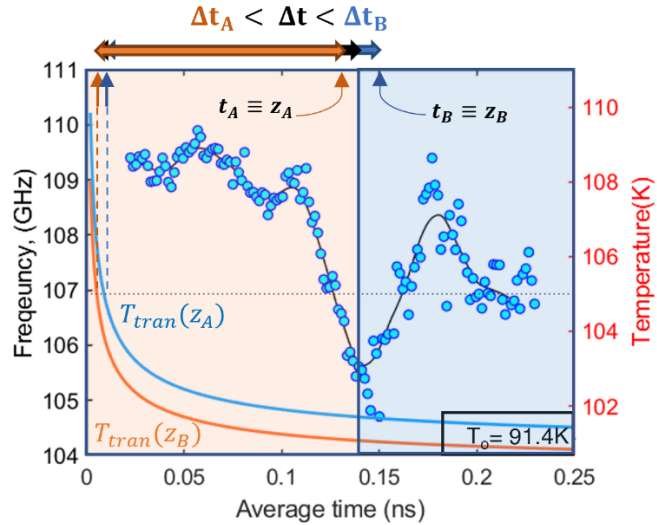

**Figure S6.** A comparative analysis between the temperature profiles of two points, $z_A$ and $z_B$, adjacent to $z_{AFD}$ illustrates the timescale of structural transition. The transient temperature profiles allow us to calculate the time difference between the arrival of the probe beam and the instance at which points A and B reach T=105K. These time differences for points A and B are defined as $\Delta t_A$ and $\Delta t_B$, respectively. While point A did not get enough time to reorient its crystal, $\Delta t_B$ was sufficiently greater than the structural delay time $\Delta t$, resulting in a tetragonal phase by the time the probe had reached.

**Thermal Model:**

In order to isolate the laser heating from thermostat temperature control, we have used a thermal wave model to quantify the thermal response of the sample at different depths. The model solves the following continuum heat diffusion equation to capture the thermal wave excited by a harmonically modulated heat source applied to the top thin metal layer[1,2].

$$\rho_s C_s \frac{\partial T}{\partial t} = k_s \nabla^2 T + Q e^{i2\pi f_m t} \tag{S1}$$

Where:

T = temperature of the sample.

$\rho_s, C_s$ = density and specific heat of the sample

$Q$ = Heat source

$f_m$ = AOM Modulation frequency

The solution to this equation is obtained by taking its Fourier transform with respect to time and then Hankel transformation with respect to the spatial coordinate. Due to axial symmetry and isotropic thermal conductivity, one can perform the Hankel transform with respect to radial coordinate r (from the pump center of incidence of the pump beam), assuming exponential heat dissipation along normal axis z[1]. The resulting expression for the thermal response takes the form of,

$$\tilde{T}(\xi) = \frac{1}{4\pi\kappa\eta} \exp(-\eta z) \tag{S2}$$

Where:

$\eta = \sqrt{\xi^2 + i2\pi f_m/D}$

$k, D$ = thermal conductivity and thermal diffusivity of the sample respectively

The next step is to model the laser pulses. The average absorbed power by the sample, $P$ has a Gaussian distribution in space with $\frac{1}{e}$ radius of beam spot size, $R$. Assuming a periodic source the beam profile can be expressed as,

$$Q(r) = \frac{P}{4\pi R^2} \exp\left(-\frac{r^2}{2R^2}\right)\left(e^{i2\pi f_m t} + 1\right) \tag{S3}$$

After applying Fourier and Hankel transformation on equation S3, one can get the expression for pump beam intensity,

$$\tilde{Q}(\xi) = P \exp\left(-\frac{\xi^2 R^2}{2}\right) \tag{S4}$$

Finally, the temperature response due to a modulated laser beam can be obtained from the product of equations S2 and S4. However, it is critical to realize the two different contributions toward the total temperature response of the sample. One is steady state temperature response ($T_{ss}$), which accounts for the average temperature rise across the sample at any given depth and is independent of time. The steady-state temperature distribution can be derived from the inverse Hankel transformation of the product of $\tilde{Q}(\xi)$ and $\tilde{T}(\xi)$,[1]

$$T_{ss}(z, f_m) = \int_0^\infty \frac{P}{4\pi\kappa} \frac{1}{\eta} e^{-\eta z} e^{-\frac{\xi^2 R^2}{2}} J_0(\xi r)\xi d\xi \tag{S5}$$

The other contribution is time-dependent which stems from the periodic nature of the pulse beam. This term essentially captures the transient temperature rise of the sample at a given location. Taking the inverse Fourier transform of equation S5 gives the time-dependent temperature distribution[3],

$$T_t(t, z, f_m, f_{rep}) = e^{-i2\pi f_m} \sum_{n=-\infty}^{\infty} e^{-i2\pi(nf_{rep})t} \int_0^\infty A(\xi, f_m + nf_{rep}) e^{-\eta z} e^{-\frac{\xi^2 R^2}{2}} \xi d\xi \tag{S6}$$

Where:

$$A(\xi, f_m + nf_{rep}) = \frac{P}{4\pi\kappa} \frac{1}{\eta}$$

$$\eta = \sqrt{\xi^2 + i2\pi(f_m + nf_{rep})/D}$$

$f_{rep}$ = Repetition rate of the laser

The combined temperature rise can be calculated by summing these two terms,

$$T(t, z, r = 0) = T_{ss}(z, f_m) + T_t(t, z, f_m, f_{rep}) \tag{S7}$$

Using this expression one can calculate the instantaneous temperature $T_{inst}(t, z = v_a t)$, which is defined as the temperature rise of the sample at any given depth of the sample at a time that corresponds to the time it takes for the probe beam to reach that depth. One can also calculate the transient temperature profile at a fixed depth which is termed as $T_{tran}(t, z_{AFD})$. It is critical to note that, the transient temperature profiles discussed in the text have contributions from both the steady-state and time-dependent terms from equations S5 and S6.